\documentclass[aps,twocolumn,prd]{revtex4}
\usepackage[dvips]{graphicx}
\usepackage{amsmath}
\usepackage{amssymb}
\newcommand{\ud}{\mathrm{d}}
\newcommand{\dirac}{\partial\llap{$\diagup$\kern-2pt}}
\newcommand{\fett}[1]{\boldsymbol{#1}}
\newcommand{\fettu}[1]{\mathbf{#1}}

\newcommand{\diag}{\mathrm{diag}}

\newcommand{\Tr}{\mathrm{Tr}}

\newcommand{\be}{\begin{equation}}
\newcommand{\ee}{\end{equation}}
\newcommand{\bea}{\begin{eqnarray}}
\newcommand{\eea}{\end{eqnarray}}
\newcommand{\bsub}{\begin{subequations}}
\newcommand{\esub}{\end{subequations}}

\begin{document}

\title{Phase diagram of dense neutral three-flavor 
quark matter}

\author{Stefan B.\ R\"uster}
\email{ruester@th.physik.uni-frankfurt.de}
\affiliation{
Institut f\"ur Theoretische Physik, J.W.\ Goethe-Universit\"at,\\
D-60054 Frankfurt am Main, Germany}

\author{Igor A.\ Shovkovy}
\email{shovkovy@th.physik.uni-frankfurt.de}
  \altaffiliation[on leave of absence from ]{%
       Bogolyubov Institute for Theoretical Physics,
       03143, Kiev, Ukraine}
\affiliation{
Institut f\"ur Theoretische Physik, J.W.\ Goethe-Universit\"at,\\
D-60054 Frankfurt am Main, Germany}

\author{Dirk H.\ Rischke}
\email{drischke@th.physik.uni-frankfurt.de}
\affiliation{
Institut f\"ur Theoretische Physik, J.W.\ Goethe-Universit\"at,\\
D-60054 Frankfurt am Main, Germany}

\date{\today}

\begin{abstract}
We study the phase diagram of dense, locally neutral three-flavor quark 
matter as a function of the strange quark mass, the quark chemical potential, 
and the temperature, employing a general nine-parameter ansatz for the
gap matrix. At zero temperature and small values of the strange quark 
mass, the ground state of matter corresponds to the color-flavor-locked 
(CFL) phase. At some critical value of the strange quark mass, this is 
replaced by the recently proposed gapless CFL (gCFL) phase. We also 
find several other phases, for instance, a metallic CFL (mCFL) phase, 
a so-called uSC phase where all colors of up quarks are paired, as 
well as the standard two-flavor color-superconducting (2SC) phase and 
the gapless 2SC (g2SC) phase.  
\end{abstract}

\maketitle

\section{introduction}

At sufficiently high densities and sufficiently low temperatures quark
matter is a color superconductor \cite{BailinLove}. This conclusion 
follows naturally from arguments similar to those employed in the case
of ordinary low-temperature superconductivity in metals and alloys
\cite{BCS}. Of course, the case of quark matter is more complicated 
because quarks, unlike electrons, come in various flavors (e.g., up, 
down, and strange) and carry non-Abelian color charges. In the last six 
years, this phenomenon was studied in detail by various authors 
\cite{cs,cfl,weak,self-e,weak-cfl,crystal,spin-1,spin-1-Meissner}. Many 
different phases were discovered, and recent studies
\cite{g2SC,GLW,var-appr,LZ,gCFL,dSC} suggest that even more new phases 
may exist. (For reviews on color superconductivity see, for example, 
Ref.~\cite{reviews}.)

In nature, the most likely place where color superconductivity may occur 
is the interior of compact stars. Therefore, it is of great importance  
to study the phases of dense matter under the conditions that are typical 
for the interior of stars. For example, one should appreciate 
that matter in the bulk of a star is neutral and $\beta$-equilibrated. 
By making use of rather general arguments, it was suggested in 
Ref.~\cite{absence2sc} that such conditions favor the
color-flavor-locked (CFL) phase and disfavor the two-flavor 
superconducting (2SC) phase. In trying to refine the validity of this 
conclusion, it was recently realized that, depending on the value of 
the constituent (medium modified) strange quark mass, the ground state 
of neutral and $\beta$-equilibrated dense quark matter may be different 
from the CFL phase \cite{g2SC,GLW,var-appr,LZ,gCFL,dSC}. In particular, 
the gapless two-flavor color-superconducting (g2SC) phase \cite{g2SC} 
is likely to be the ground state in the case of a large strange quark 
mass. On the other hand, in the case of a moderately large strange quark 
mass, the regular and gapless color-flavor-locked (gCFL) phases \cite{gCFL} 
are favored. At nonzero temperature, $T\neq 0$, some other phases were 
proposed as well \cite{dSC}.

In this paper, we study the phase diagram of dense neutral three-flavor
quark matter as a function of the strange quark mass at zero and finite 
temperature. In order to allow for the most general ground state, we 
employ a nine-parameter ansatz for the gap function. The effects of the 
strange quark mass are incorporated in our model by a shift of the chemical
potential of strange quarks, $\mu^{i}_{s} \to \mu^{i}_s - m_s^2/(2\mu)$
where $i=r,g,b$ is the color index, $m_s$ is the strange quark mass, 
and $\mu$ is the quark chemical potential. This shift reflects the 
reduction of the Fermi momenta of strange quarks due to their mass. 
Such an approach is certainly reliable at small values of the strange 
quark mass. We assume that it is also qualitatively correct at large 
values of the strange quark mass. 

We note that the analysis of this paper is restricted to locally neutral 
phases only. This automatically excludes, for example, mixed \cite{mix} 
and crystalline \cite{cryst} phases. Also, in the mean field 
approximation utilized here, we cannot get any phases with 
meson condensates \cite{condensate}.

This paper is organized as follows. In Sec.~\ref{eos} we present the
details of our approach which is based on the Cornwall-Jackiw-Tomboulis 
(CJT) formalism \cite{CJT}. There, we also derive the gap equations and 
the neutrality conditions, and obtain an expression for the pressure.
In the next two sections, the gap equations and the neutrality 
conditions are studied by using numerical methods. The results at 
zero temperature are presented and discussed in Sec.~\ref{results}. 
There, the appearance of the CFL and gCFL phases is established in 
the case of small and moderately large values of the strange quark mass, 
respectively. The results at nonzero temperature are discussed in 
Sec.~\ref{results-T}. Also, the phase diagram of quark matter in the
$T$--$m_s^2/\mu$ plane, as well as in the $T$--$\mu$ plane is presented. 
Finally, Sec.~\ref{conclusions} concludes 
this paper with a summary of the results.

Our units are $\hbar=c=k_B=1$. The metric tensor is $g_{\mu\nu}
=\diag\left(1,-1,-1,-1\right)$. Four-vectors are denoted by capital Latin 
letters, e.g., $K^\mu=(k^ 0,\fettu{k})$ where $\fettu{k}$ is a three-vector 
with absolute value $k=|\fettu{k}|$ and direction $\hat{\fettu{k}}=
\fettu{k}/k$. We use the imaginary-time 
formalism, i.e., the space-time integration is defined as $\int_X=
\int_0^{1/T}\ud\tau\int_V\ud^3\fettu{x}$, where $\tau$ is the Euclidean 
time coordinate and $V$ the three-volume of the
system. Energy-momentum sums are defined as follows: $T/V\sum_K
=T\sum_n\int\ud^3\fettu{k}/(2\pi)^3$ where the sum runs over the fermionic 
Matsubara frequencies $\omega_n=(2n+1)\pi T\equiv ik_0$.

\section{Model and formalism}
\label{eos}

In this paper, we follow closely the approach of Ref.~\cite{Paper1} 
which is based on the CJT formalism \cite{CJT}. We generalize the model of 
Ref.~\cite{Paper1} in the way that strange quarks may now also participate 
in forming Cooper pairs. The quark spinor field has the following 
color-flavor structure:
\be
\label{spinorbasis}
\psi=\left(
\begin{array}{c}
   \psi_u^r \\
   \psi_d^r \\
   \psi_s^r \\
   \psi_u^g \\
   \psi_d^g \\
   \psi_s^g \\
   \psi_u^b \\
   \psi_d^b \\
   \psi_s^b
\end{array}
\right)\, .
\ee
The Dirac conjugate spinor is defined as $\bar\psi=\psi^\dagger\gamma_0$.
In the treatment of superconducting systems it is advantageous to double
the fermionic degrees of freedom by introducing the so-called Nambu-Gorkov 
spinors
\be
\label{Psi}
\bar\Psi=\left( \bar\psi,\bar\psi_C \right) \; , \qquad
\Psi = \left(
\begin{array}{c}
  \psi \\
  \psi_C
\end{array}
\right) \; ,
\ee
where $\psi_C=C\bar\psi^T$ is the charge-conjugate spinor, and $C$ is the
charge-conjugation matrix.

In this paper, as in Ref.~\cite{Paper1}, we shall approximate the 
gluon-exchange interaction between the quarks by a point-like 
current-current interaction. Physically, this is equivalent to a 
model with heavy, non-dynamical gluons. In many ways, such an 
approximation is analogous to the nonrenormalizable Fermi theory 
of weak interactions that was used before the SU(2)$\times$U(1) 
gauge theory of electro-weak forces was developed \cite{EW}. Also, 
in the mean-field approximation that we use below, our treatment 
will be similar to that in 
Refs.~\cite{cs,cfl,g2SC,GLW,LZ,gCFL,4fermi,4fermi-m,NJL-dense} 
which are based on Nambu--Jona-Lasinio (NJL) \cite{NJL} type models.

In the approximation with a point-like interaction, the CJT effective 
action simplifies. It has the following general form:
\be
\label{Gamma}
\Gamma \left[ \mathcal{S} \right]
=\frac12 \Tr \ln \mathcal{S}^{-1}
+\frac12 \Tr \left( S_0^{-1} \mathcal{S} - 1 \right)
+\Gamma_2 \left[ \mathcal{S} \right] \; .
\ee
Here the traces run over space-time, Nambu-Gorkov, color, flavor, and 
Dirac indices. The factor $1/2$ in front of the fermionic one-loop terms 
compensates the doubling of the degrees of freedom in the Nambu-Gorkov 
basis. In the above effective action, $\mathcal{S}$ 
denotes the full quark propagator.

In order to understand the structure of the full quark propagator 
$\mathcal{S}$, let us first discuss the structure of the free quark 
propagator $S_0$ in the Nambu-Gorkov basis. The inverse, $S_0^{-1}$, 
is given by the following matrix:
\be
S_0^{-1} = 
\left(
\begin{array}{cc}
[ G_0^+ ]^{-1} & 0 \\
0 & [ G_0^- ]^{-1}
\end{array}
\right) \; ,
\ee
where
\be
[ G_0^\pm ]^{-1} = \gamma^\mu K_\mu \pm \hat \mu \gamma_0
\ee
are the inverse Dirac propagators for massless quarks and charge-conjugate 
quarks,
respectively. At sufficiently large quark chemical potential, there is 
no need to take into account the small
masses of the light up- and down-quarks. It is easy to understand that 
the dynamical effect of such masses around the quark Fermi surfaces is 
negligible. Of course, the situation with the strange quark is different
because its mass is not very small as compared to $\mu$. 
It appears, however, that the most 
important effect of a nonzero strange quark mass is a shift of the 
strange quark chemical potential due to the reduction of the Fermi
momentum, $\mu_{s}^{i}\to \mu_s^i - m_s^2/(2\mu)$. 
(Strictly speaking, it is $\mu_s^i$ rather than $\mu$ that should 
appear in the denominator. Quantitatively, however, this does not make 
a big difference.) For simplicity, this is the only mass effect that we 
include in our analysis below. Note that nonzero quark masses were
properly accounted for in Refs.~\cite{4fermi-m,NJL-dense,TF}.

The quark chemical potential matrix $\hat\mu$ in color-flavor 
space is defined as 
\be
\hat\mu = \diag \left( \mu_u^r, \mu_d^r, \mu_s^r, \mu_u^g,
\mu_d^g, \mu_s^g ,\mu_u^b, \mu_d^b, \mu_s^b \right).
\label{hat-mu}
\ee
In realistic systems, such as bulk matter inside a star, some of the
$\mu_f^i$ should be related. This is because, in chemical equilibrium,
one can introduce only as many independent chemical potentials as there
are different conserved charges in the system. In quark matter, the 
chemical potentials $\mu_f^i$ can be defined in terms of the quark 
chemical potential $\mu$ ($\mu\equiv \mu_B/3$, where $\mu_B$ is the 
baryon chemical potential), the chemical potential $\mu_Q$ for the 
electrical charge, and the two chemical potentials, $\mu_3$ and $\mu_8$, 
for color charge,
\be
\mu_f^i = \mu + \mu_Q Q_f + \mu_3 T_3^i + \mu_8 T_8^i.
\label{mu-f-i}
\ee
Thus, there are only four out of nine chemical potentials in 
Eq.~(\ref{hat-mu}) that are independent in the case of dense three-flavor 
quark matter in chemical equilibrium. In (locally) neutral matter, as we
shall see below, only one of them will remain independent.

In the framework of the CJT formalism, the thermodynamic potential 
of quark matter is proportional to the CJT effective action at its 
stationary point, determined by the solution of the following 
equation:
\be
\frac{ \delta \Gamma }{ \delta \mathcal{S} }
=0.
\label{st-p}
\ee
This is nothing but the Dyson-Schwinger equation for the quark 
propagator,
\be
\label{Shm1}
\mathcal{S}^{-1} = S_0^{-1} + \Sigma \; , 
\ee
where
\be \Sigma\equiv 
2\frac{\delta\Gamma_2\left[ \mathcal{S}\right]}{\delta \mathcal{S}}
\ee
is the quark self-energy. The functional $\Gamma_2$ is the sum of all 
two-particle 
irreducible (2PI) diagrams. Unfortunately, it is impossible to evaluate 
all 2PI diagrams exactly. Nevertheless, the advantage of the CJT effective 
action (\ref{Gamma}) is that a truncation of the sum $\Gamma_2$ 
at a finite number of terms still provides a well-defined many-body 
approximation. In this study, we only include the sunset-type diagram 
shown in Fig.~1 of Ref.~\cite{Paper1}, which becomes a double-bubble 
diagram in the case of a local, instantaneous interaction. It is easy 
to check that this leads to the following expression for the self-energy:
\be
\label{gapeqn}
\Sigma \left( K \right) = - g^2 \frac{T}{V} \sum_Q \Gamma_a^\mu 
\mathcal{S} \left( Q \right) \Gamma_b^\nu D_{\mu\nu}^{ab} \; .
\ee
Local, instantaneous gluon exchange is parametrized by a
propagator of the form
\be
D_{\mu\nu}^{ab} \equiv - \delta^{ab} \frac{ g_{\mu\nu} }{\Lambda^2 } .
\label{gl-pro}
\ee
In Eq.~(\ref{gapeqn}), we introduced the Nambu-Gorkov vertex,
\be
\Gamma_a^\mu = \left(
\begin{array}{cc}
\gamma^\mu T_a & 0 \\
0 & -\gamma^\mu T_a^T
\end{array}
\right).
\ee
Equation (\ref{gapeqn}) is a self-consistency equation for $\Sigma$,
commonly called the ``gap equation''.

In color-superconducting systems, the quark self-energy has the 
form
\be
\Sigma =
\left(
\begin{array}{ll}
\, 0 & \Phi^- \\
\Phi^+ & \, 0
\end{array}
\right).
\label{sig}
\ee
The regular (diagonal) part of the self-energy is 
neglected here. While it plays an important role in the 
dynamics of chiral symmetry breaking, it is essentially irrelevant 
for color superconductivity in dense quark matter. (The effect of the 
diagonal part of the self-energy was studied, for example, in 
Ref.~\cite{self-e}.) The off-diagonal elements
$\Phi^{-}$ and $\Phi^{+}$ in Eq.~(\ref{sig}) are matrices in Dirac, 
color and flavor space. They are related by the requirement that the
action is real, 
$\Phi^{-}\equiv\gamma^0 \left(\Phi^{+}\right)^{\dagger}\gamma^0$. 

With Eq.~(\ref{sig}) one may invert Eq.~(\ref{Shm1}) to obtain the 
full quark propagator,
\be
\mathcal{S} = \left(
\begin{array}{cc}
G^+ & \Xi^- \\
\Xi^+ & G^-
\end{array}
\right) \; ,
\ee
where the diagonal and off-diagonal elements are
\be
G^\pm = \left\{ [ G_0^\pm ]^{-1} - \Phi^\mp G_0^\mp \Phi^\pm \right\}^{-1}
\ee
and
\be
\Xi^\pm = - G_0^\mp \Phi^\pm G^\pm ,
\ee
respectively. 

In the following, we use the notation $\Gamma^{*}$ for the CJT 
effective action evaluated at the stationary point. In the 
approximation used here, it takes the following form:
\be
\label{Gamma2}
\Gamma^{*} 
=\frac12 \Tr \ln \mathcal{S}^{-1}
-\frac14 \Tr \left( \Sigma \mathcal{S} \right) \; ,
\ee
In this 
paper, we utilize the following nine-parameter ansatz for the 
gap matrix:
\begin{widetext}
\be
\label{Phi}
\Phi^\pm =
\left(
\begin{array}{ccccccccc}
[ \Delta_{uu}^{rr} ]^\pm & 0 & 0 & 0 & [ \Delta_{ud}^{rg} ]^\pm & 0 & 0 &
0 & [ \Delta_{us}^{rb} ]^\pm \\
0 & 0 & 0 & [ \Delta_{du}^{rg} ]^\pm & 0 & 0 & 0 & 0 & 0 \\
0 & 0 & 0 & 0 & 0 & 0 & [ \Delta_{su}^{rb} ]^\pm & 0 & 0 \\
0 & [ \Delta_{du}^{rg} ]^\pm & 0 & 0 & 0 & 0 & 0 & 0 & 0 \\
{[ \Delta_{ud}^{rg} ]}^\pm & 0 & 0 & 0 & [ \Delta_{dd}^{gg} ]^\pm & 0 & 0
& 0 & [ \Delta_{ds}^{gb} ]^\pm \\
0 & 0 & 0 & 0 & 0 & 0 & 0 & [ \Delta_{sd}^{gb} ]^\pm & 0 \\
0 & 0 & [ \Delta_{su}^{rb} ]^\pm & 0 & 0 & 0 & 0 & 0 & 0 \\
0 & 0 & 0 & 0 & 0 & [ \Delta_{sd}^{gb} ]^\pm & 0 & 0 & 0 \\
{[ \Delta_{us}^{rb} ]}^\pm & 0 & 0 & 0 & [ \Delta_{ds}^{gb} ]^\pm & 0 & 0
& 0 & [ \Delta_{ss}^{bb} ]^\pm
\end{array}
\right) \; . \\
\ee
\end{widetext}
The explicit Dirac structure of the matrix elements is
\bsub
\bea
{[ \Delta_{ff'}^{ii'} ]}^+ \left( K \right) &=& \sum_{c,e=\pm}
{\phi_c^e}_{ff'}^{ii'} \left( K \right) \mathcal{P}_c^e 
(\fettu{k}) \; , \\
{[ \Delta_{ff'}^{ii'} ]}^- \left( K \right) &=& \sum_{c,e=\pm}
{\phi_c^e}_{ff'}^{ii'} \left( K \right) \mathcal{P}_{-c}^{-e} 
(\fettu{k}) \; .
\eea
\label{real-phi}
\esub
In this representation, ${\phi_c^e}_{ff'}^{ii'}$ are real-valued gap 
parameters, while
\be
\mathcal{P}_c^e ( \fettu{k}) = \textstyle \frac14
\displaystyle \, ( 1 + c \gamma_5 )( 1 + e \gamma_0 \fett{\gamma} \cdot
\hat{\fettu{k}} )
\ee
are the energy-chirality projectors.

\begin{figure}[ht]
  \begin{center}
    \includegraphics[width=0.45\textwidth]{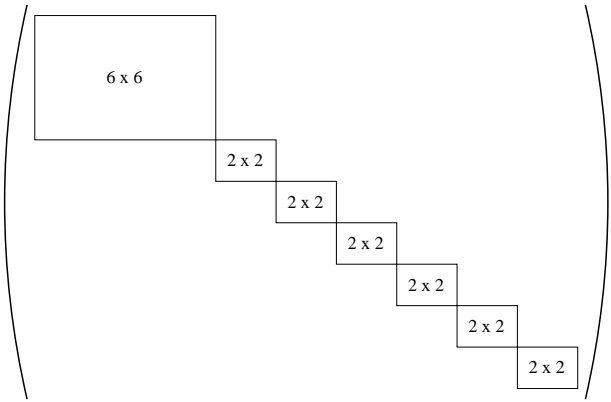}
    \caption{The block-diagonal structure of the inverse quark 
             propagator in color-flavor and Nambu-Gorkov space.}
    \label{block}
  \end{center}
\end{figure}

In order to calculate the first part of the effective action in 
Eq.~(\ref{Gamma2}), we transform the inverse quark propagator 
(\ref{Shm1}) into a block-diagonal form which is schematically 
shown in Fig.~\ref{block}. This is easily achieved by changing 
the order of rows and columns in color, flavor, and Nambu-Gorkov 
space. The energy-chirality projectors simplify the calculation of 
the Dirac traces. After performing these traces, we use 
the matrix relation $\Tr \ln A = \sum_K\ln \left(\det A\right)$
to calculate the one-loop contribution of the quarks to $\Gamma^{*}$.

We use the Gauss elimination procedure to reduce 
the $6 \times 6$-block to a $3 \times 3$-block and all the
$2 \times 2$-blocks to $1 \times 1$-blocks in color-flavor space. 
The determinant of the $3 \times 3$-block is computed analytically
by using, for example, Mathematica or Maple.

The most complicated expression arises from the determinant
of the $3 \times 3$-block. Schematically, it has the following 
form:
\be
\det\left(3 \times 3\right)_e = k_0^6 + b k_0^4 + c k_0^2 + d. 
\label{det33}
\ee
Here, the coefficients $b$, $c$, and $d$ are rather complicated 
functions of the quark momentum $k$, three chemical potentials 
($\mu_u^r$, $\mu_d^g$, and $\mu_s^b$) and six gap parameters 
($\phi_{uu}^{rr}$, $\phi_{ud}^{rg}$, $\phi_{us}^{rb}$, 
$\phi_{dd}^{gg}$, $\phi_{ds}^{gb}$, and $\phi_{ss}^{bb}$). 
This determinant can always be factorized as follows: 
$\det\left(3 \times 3\right)_e\equiv [k_0^2-(\tilde\epsilon^e_1)^2]
[k_0^2-(\tilde\epsilon^e_2)^2] [k_0^2-(\tilde\epsilon^e_3)^2]$.
As is clear, the functions $\tilde\epsilon^e_1$, $\tilde\epsilon^e_2$ 
and $\tilde\epsilon^e_3$ determine the dispersion relations of 
the three quasiparticles described by the $3 \times 3$-block of the quark 
propagator. In order to get their explicit expressions, one solves 
the cubic equation $\det\left(3 \times 3\right)_e=0$ for $\xi=k_0^2$. 
By making use of Cardano's formula, the solutions can be presented 
in an analytical form that we use later in the numerical calculations. 
Because of their very complicated nature we refrain from presenting
them explicitly. 

The dispersion relations of the quasiparticles that 
correspond to the $1 \times 1$-blocks are easy to derive. 
They are given in terms of the same function, 
$\epsilon_\fettu{k}^e(\mu,\phi) = [(k-e\mu)^2 + |\phi|^2]^{1/2}$,
that also appears in the 2SC phase.

Once the quasiparticle dispersion relations are established, the 
calculation of the first term in the CJT action, i.e., 
$\sum_K\ln \left(\det\mathcal{S}^{-1}\right) \equiv 
\sum_{j}\sum_K \ln \left[(\omega_n+i\mu_j)^2+\epsilon_j^2\right]$,
reduces to a sum of several standard contributions of the following 
type \cite{Kapusta}:
\bea
\sum_K \ln\left(\frac{(\omega_n+i\mu_j)^2+\epsilon_j^2}{T^2}\right) =
V\int\frac{p^2dp}{2\pi^2}\Bigg[\frac{\epsilon_j}{T} \qquad
&&\nonumber\\
+\ln\left(1+\mbox{e}^{-(\epsilon_j-\mu_j)/T} \right) 
+\ln\left(1+\mbox{e}^{-(\epsilon_j+\mu_j)/T} \right) 
\Bigg].&&
\eea
In order to simplify the second term in the effective potential in 
Eq.~(\ref{Gamma2}), we use the gap equation (\ref{gapeqn}).
Then, we arrive at the following simple result:
\be
- \frac14 \Tr \left( \Sigma S \right) = \frac34 \frac{ \Lambda^2 }{ g^2 }
\frac{ V }{ T } \, \Tr \left[ \hat \Phi \left( \hat \Phi + 3 \sum_a T_a^T
\hat\Phi T_a \right) \right] \; , \\
\ee
where the new notation $\hat{\Phi}$ was introduced to represent only 
the color-flavor part of the gap matrix $\Phi^+$, given in Eq.~(\ref{Phi}).
In this connection, it should be noted that the gap parameters 
${\phi_c^e}_{ff'}^{ii'}$ are independent of the chirality and energy 
projections. This is a consequence of the gap equation (\ref{gapeqn}).

Finally, by combining all contributions to Eq.~(\ref{Gamma2}), we derive 
the result for the pressure $p\equiv\textstyle \frac{T}{V} \Gamma^*$,
\begin{widetext}
\bea
p &=& \frac{T}{\pi^2} \sum_{l}
\int_0^\infty \ud k \, k^2 \left\{ 
  \ln \left[ 1 + \exp\left(-\frac{E_l-\mu_l}{T}\right)\right]
+ \ln \left[ 1 + \exp\left(-\frac{E_l+\mu_l}{T}\right)\right]\right\}
 \nonumber \\
&+& \frac34 \frac{\Lambda^2}{g^2} 
  \sum_{i=1}^3 \left(\phi_{i}^2+\varphi_{i}^2
                   +6 \phi_{i} \varphi_{i} 
                   +2 \sigma_{i}^2 \right) 
+\frac{1}{2\pi^2} \sum_{i=1}^3 \sum_{e=\pm} \int_0^\kappa \ud k \, k^2 
\left\{ \tilde\epsilon_i^e - k 
+ 2T \ln \left[ 1 + \exp\left(- \frac{\tilde\epsilon_i^e}{T}\right) \right] 
\right\} \nonumber \\
&+& \frac{1}{\pi^2} \sum_{i=1}^3 \sum_{e=\pm} \int_0^\kappa \ud k \, k^2 
\bigg\{ \epsilon_\fettu{k}^e \left( \bar\mu_i, \phi_i \right) - k
+ T \ln \left[ 1 + \exp \left( - \, \frac{ \epsilon_\fettu{k}^e 
\left( \bar\mu_i, \phi_i \right) + \delta\mu_i }{ T } \right) \right]  
\nonumber \\
&&\hspace*{2.2in}
+ T \ln \left[ 1 + \exp \left( - \, \frac{ \epsilon_\fettu{k}^e 
\left( \bar\mu_i, \phi_i \right) - \delta\mu_i }{ T } \right) \right] 
\bigg\} \; . 
\label{pressure}
\eea
\end{widetext}
The first term in this expression is the contribution of leptons
(i.e., $l=e^{-},\mu^{-}$). In principle, the contribution of 
neutrinos should be added as well. In this paper, however, their
contribution is neglected. This is a good approximation for 
compact stars after deleptonization.
The dispersion relations of the leptons
are given by $E_l=\sqrt{k^2 + m_l^2}$ with $m_{e^{-}}\approx 0.511$ MeV
and $m_{\mu^{-}}\approx 105.66$ MeV. Note that the vacuum 
contribution was subtracted in Eq.~(\ref{pressure}), in order to improve 
the convergence of the momentum integrals. In various applications, 
a bag constant could be added if necessary.
In Eq.~(\ref{pressure}), $\phi_i$ are the three independent gap parameters 
that appear in the $2 \times 2$-blocks in Fig.~\ref{block}, while 
$\varphi_i$ and $\sigma_i$ appear in the $6 \times 6$-block. They are 
defined by 
\bsub
\bea
\phi_1 \equiv \phi_{sd}^{gb}, 
\qquad 
\varphi_1 \equiv \phi_{ds}^{gb} ,
\quad\mbox{and}\quad 
\sigma_1 \equiv \phi_{uu}^{rr},\\
\phi_2 \equiv \phi_{su}^{rb},  
\qquad 
\varphi_2 \equiv \phi_{us}^{rb} ,
\quad\mbox{and}\quad 
\sigma_2 \equiv \phi_{dd}^{gg},\\
\phi_3 \equiv \phi_{du}^{rg},
\qquad 
\varphi_3 \equiv \phi_{ud}^{rg} ,
\quad\mbox{and}\quad 
\sigma_3 \equiv \phi_{ss}^{bb} . 
\eea
\esub
In accordance with the fact that the gap parameters are independent 
of the energy and chirality, the corresponding indices were dropped. 
Similarly, $\bar\mu_i$ and $\delta\mu_i$ ($i=1,2,3$) are the 
average values and the differences of various pairs of 
chemical potentials that come from the three different 
$2 \times 2$-blocks, 
\bsub
\bea
\bar\mu_1 \equiv \frac12 ( \mu_s^g + \mu_d^b ),
\quad\mbox{and}\quad
\delta\mu_1 \equiv \frac12 ( \mu_s^g - \mu_d^b ),&&\\
\bar\mu_2 \equiv \frac12 ( \mu_s^r + \mu_u^b ),
\quad\mbox{and}\quad 
\delta\mu_2 \equiv \frac12 ( \mu_s^r - \mu_u^b ),&&\\
\bar\mu_3 \equiv \frac12 ( \mu_d^r + \mu_u^g ),
\quad\mbox{and}\quad
\delta\mu_3 \equiv \frac12 ( \mu_d^r - \mu_u^g ).&&
\eea
\esub
In order to render the integrals in the expression for the pressure 
finite, we introduced a three-momentum cutoff $\kappa$. In QCD with 
dynamical gluons, of course, such a cutoff would not be
necessary. Here, however, we use a model with a local 
current-current interaction which is nonrenormalizable.

In the case of the ansatz (\ref{Phi}) for the gap function, there 
are nine coupled gap equations (corresponding to the total number 
of gap parameters) that follow from Eq.~(\ref{gapeqn}). The 
explicit form of these equations is very complicated and not very
informative by itself. Because of this, they will not be displayed 
here. In the next two sections, we solve these gap equations by 
numerical methods. 

As was mentioned in the Introduction, matter in the bulk of a compact
star should satisfy the conditions of charge neutrality and 
$\beta$-equilibrium. The latter is automatically satisfied after 
choosing the chemical potentials as in Eq.~(\ref{mu-f-i}). The 
neutrality conditions read
\bsub
\label{neutrality}
\bea
n_Q &\equiv& \frac{ \partial p }{\partial \mu_Q} = 0 \; , \\
n_3 &\equiv& \frac{ \partial p }{\partial \mu_3} = 0 \; , \\
n_8 &\equiv& \frac{ \partial p }{\partial \mu_8} = 0 \; .
\eea
\esub
After solving these equations, one is left with only one chemical potential
(out of four) that remains independent. It is most convenient to 
keep the quark chemical potential $\mu$ as a free parameter, and determine
$\mu_Q$, $\mu_3$ and $\mu_8$ from Eqs.~(\ref{neutrality}).
This is done numerically.

\section{Results at zero temperature}
\label{results}

In this section we focus on three-flavor quark matter 
at zero temperature. It is clear that, for small and moderate values 
of the strange quark mass, the ground state of neutral quark matter
should correspond to either the regular (gapped) CFL phase \cite{cfl} 
or the gapless CFL phase \cite{gCFL}. At very large strange quark mass
and/or relatively weak coupling, the ground state can also be either 
a regular (gapped) or gapless 2SC color superconductor \cite{g2SC}. 

Before proceeding to the results, let us specify the parameters
of the model. The strength of the diquark coupling and 
the value of the cutoff in the momentum integrals are fixed as follows: 
\bsub
\label{modelparameters}
\bea
g^2 / \Lambda^2 &=& 45.1467 \ \mathrm{GeV}^{-2} \; , \\
\kappa &=& 0.6533 \ \mathrm{GeV} \; .
\eea
\esub
In order to see how the phase structure of neutral three-flavor 
quark matter changes with the mass of the strange quark $m_s$, we 
solve a coupled set of twelve equations, i.e., nine gap equations and
three neutrality conditions, given in Eq.~(\ref{neutrality}), for various
values of $m_s$, keeping the quark chemical potential fixed. In this 
calculation we take $\mu=500$ MeV. The results for the absolute values
of the gap parameters and the chemical potentials $\mu_Q$, $\mu_3$, 
and $\mu_8$ are 
shown in Figs.~\ref{phi2ms} and \ref{ch2ms}, respectively. Note that, 
strictly speaking, the gap 
parameters do not coincide with the actual values of the gaps in the 
quasiparticle spectra. In the case of the CFL phase, for example, there 
is a degenerate octet of quasiparticles with a gap 
$\phi_{\rm octet}= |\phi_{1}|$ 
and a singlet state with a gap 
$\phi_{\rm singlet}=3\varphi_{1}-|\phi_{1}|$. 
In the CFL phase, $\phi_{1}=\phi_{2}=\phi_{3}<0$,
$\varphi_{1}=\varphi_{2}=\varphi_{3}>0$, and 
$\sigma_{1}=\sigma_{2}=\sigma_{3}>0$. 
Also, in the CFL phase, the following relation between the gap 
parameters is satisfied: 
$\sigma_{i}=\varphi_{i}-|\phi_{i}|\equiv 2\phi_{(6,6)}$, $i=1,2,3$,
where $\phi_{(6,6)}$ is the sextet-sextet gap in the 
notation of Ref.~\cite{weak-cfl}.

The results in Figs.~\ref{phi2ms} and \ref{ch2ms} extend
the results of Ref.~\cite{gCFL} by considering a more general
ansatz for the gap matrix that takes into account, in particular, 
the pairing in the symmetric sextet-sextet channel. 
The effect of including pairing in the 
symmetric channel is a splitting between the pairs of gaps 
($|\phi_{1}|$, $\varphi_{1}$), 
($|\phi_{2}|$, $\varphi_{2}$) and 
($|\phi_{3}|$, $\varphi_{3}$) that 
is also reflected in the change of the quasiparticle spectra.
Of course, in agreement with the general arguments of 
Refs.~\cite{cfl,weak-cfl}, the symmetric sextet-sextet gaps 
are rather small, see Fig.~\ref{phi2ms}~(c). This, in turn, 
explains the fact why the splittings between the above mentioned
pairs of gap parameters are not very large [compare the results 
in Figs.~\ref{phi2ms}~(a) and (b)].

In this study, we confirm that the phase transition from the CFL phase to 
the gCFL phase happens at a critical value of the parameter $m_s^2/\mu$ 
that is in good agreement with the simple estimate of Ref.~\cite{gCFL},
\be
\frac{m_s^2}{\mu} \simeq 2\Delta\sim 190 \mbox{~MeV}.
\ee
The qualitative results for the chemical potentials $\mu_Q$, $\mu_3$, 
and $\mu_8$ in Fig.~\ref{ch2ms} are in agreement with the corresponding 
results obtained in Ref.~\cite{gCFL} as well. In the CFL phase, in our 
notation the color chemical potential $\mu_8$ fulfils the identity 
$\mu_8= -m_s^2/(\sqrt{3}\mu)$. In addition, while the CFL phase 
requires no electrons to remain neutral, the pairing in the gCFL phase 
is distorted and a finite density of electrons appears. This is seen 
directly from the dependence of the electrical chemical potential $\mu_Q$ 
in Fig.~\ref{ch2ms}, which becomes nonzero only in the gCFL phase. This 
observation led the authors of Ref.~\cite{gCFL} to the conclusion that
the phase transition between the CFL and gCFL phase is an insulator-metal
phase transition, and that the value of the electron density is a 
convenient order parameter in the description of such a transition. 
In fact, one could also choose one of the differences between number 
densities of mutually paired quarks as an alternative choice for the 
order parameter \cite{g2SC}. In either case, there does not seem to 
exist any continuous symmetry that is associated with such an order 
parameter. To complete the discussion of the chemical potentials, we 
add that the other color chemical potential, $\mu_3$, is zero only in 
the CFL phase at $T=0$. 

The effects of a nonzero strange quark mass on the phase 
structure of neutral strange quark matter could be viewed from 
a different standpoint that, in application to stars, may look 
more natural. This is the case where the dependence on the quark
chemical potential is studied at a fixed value of $m_s$. The 
corresponding numerical results are shown in Fig.~\ref{phi2mu}.
(Note once again that $\phi_{i}$ with $i=1,2,3$ have negative 
values, and we always plot their absolute values.) In this 
particular calculation we choose $m_s=300$ MeV. 

At large values of the quark chemical 
potential [$\mu\agt m_s^2/(2\Delta) \sim 475$ MeV which is similar 
to the small strange quark mass limit considered before], the ground 
state of quark matter is the CFL phase. At smaller values of the 
chemical potential, the ground state of dense matter is the gCFL 
phase. In this case, there are nine gap parameters all of which are 
different from each other. One could also check that the density 
of quarks that pair are not equal in the gCFL phase. This can be 
seen from Fig.~\ref{n2mu} where all nine quark number densities are
plotted for the same value of the strange quark mass, $m_s=300$ 
MeV. Only $n_d^r = n_u^g$ and $n_s^r \approx n_u^b$, all 
other quark number densities are different from each other. This
agrees with the general criterion of the appearance of gapless phases 
at $T=0$ that was proposed in Ref.~\cite{g2SC} in the case of two-flavor 
quark matter. In the ordinary CFL phase, in contrast, one finds that 
$n_u^r = n_d^g = n_s^b$, and $n_d^r = n_s^r = n_u^g = n_s^g 
= n_u^b = n_d^b$.

In order to see that the gCFL phase indeed describes a gapless 
superconductor, it is necessary to show that the dispersion relations 
of quasiparticles contain gapless excitations. In Fig.~\ref{disp}, the 
dispersion relations of all nine quasiparticles are plotted. 
The dispersion relations for the corresponding anti-particles 
are not shown. The 
dispersion relations that result from the $6 \times 6$ color-flavor 
block of the quark propagator are labelled by $\tilde \epsilon_i$ 
in Fig.~\ref{disp} (see the discussion and Fig.~\ref{block} in 
Sec.~\ref{eos}). The remaining dispersion relations correspond to 
quasiparticles that are described by the $2 \times 2$-blocks of the 
quark propagator, given by 
$k_0=\epsilon_\fettu{k}^e (\bar\mu_i, \phi_i) \pm \delta\mu_i$. 
Qualitatively, these are the same as the dispersion relations that 
appear in the 2SC/g2SC phase in Ref.~\cite{g2SC}. This is not a 
coincidence because the corresponding $2 \times 2$ 
color-flavor blocks of the propagator have the same structure as 
in the case of the 2SC/g2SC phase. 

From Fig.~\ref{disp} we see that there is indeed a gapless mode in the 
``green-strange--blue-down'' sector. This is the same that was found 
in Ref.~\cite{gCFL}. Note also that, in agreement with Ref.~\cite{gCFL},
the ``red-strange--blue-up'' quasiparticle has a dispersion relation 
that is nearly quadratic, $k_{0su}^{\phantom{0}rb}\simeq |k-k^{*}|^2$ 
with $k^{*}\approx 400$ MeV for a given choice of parameters, see 
Fig.~\ref{disp}~(c). The nearly quadratic dispersion relation
resembles the situation at the transition between the 2SC phase, 
where $n_u^r = n_d^g$, and the gapless 2SC phase, where $n_u^r$ and 
$n_d^g$ are different. This explains the approximate equality 
$n_s^r \approx n_u^b$ mentioned above.

\section{Results at nonzero temperature}
\label{results-T}

In this section, we present the results for the phase structure  
of dense neutral three-flavor quark matter in the plane of temperature
and $m_s^2/\mu$, as well as in the plane of temperature and quark chemical
potential.

Let us start with the discussion of the temperature dependence of the 
gap parameters in the two qualitatively different cases of small and large 
values of the strange quark mass. As we saw in the previous section, the 
zero-temperature properties of neutral quark matter were very different
in these two limits.

The results for the temperature dependence of the gap parameters 
are shown in Figs.~\ref{m_s-small}, \ref{m_s-small-zoom} and 
\ref{m_s-large} for two different values of the strange quark mass 
that represent the two qualitatively different regimes. In the case 
of a small strange quark mass (i.e., the case of $m_s^2/\mu=80$ MeV 
which is shown in Figs.~\ref{m_s-small} and \ref{m_s-small-zoom}), 
the zero-temperature limit corresponds to the CFL phase. This is 
seen from the fact that the three different gaps shown in every panel
of Fig.~\ref{m_s-small} merge as $T\to 0$. At nonzero temperature, 
on the other hand, the gap parameters are not the same. This suggests 
that, similar to the zero-temperature case of Figs.~\ref{phi2ms}
and \ref{phi2mu}, a phase transition to the gCFL phase happens 
at some nonzero temperature. However, we shall show below that there 
is no phase transition between the CFL and gCFL phases at {\em any\/}
nonzero temperature. Instead, there is an insulator-metal crossover 
transition between the CFL phase and a so-called {\em metallic\/} CFL 
(mCFL) phase. At this point, all quasiparticles are still gapped.
At some higher temperature, the mCFL phase is replaced by the gCFL 
phase.

If the temperature is increased even further, there are three 
consecutive phase transitions. These correspond to the three 
phase transitions predicted in Ref.~\cite{dSC} in the limit 
of a small strange quark mass. In order to resolve these, we 
show a close-up of the near-critical region of Fig.~\ref{m_s-small} 
in Fig.~\ref{m_s-small-zoom}. The three transitions that we observe 
are the following: 
(i) transition from the gCFL phase to the so-called uSC phase; 
(ii) transition from the uSC phase to the 2SC phase; 
(ii) transition from the 2SC phase to the normal quark phase.
Here, the notation uSC (dSC) stands for superconductivity in which 
all three colors of up (down) quark flavor participate in diquark 
pairing \cite{dSC}. Our results differ from those of Ref.~\cite{dSC} 
in that the dSC phase is replaced by the uSC phase. The reason is that,
in our case, the first gaps  which vanish with increasing temperature 
are $\phi_1$ and $\varphi_1$, see Fig.~\ref{m_s-small-zoom}, while 
in their case $\Delta_2$ (corresponding to our $\phi_2$ and $\varphi_2$) 
disappears first. Although we use the same terms for the phases that 
were introduced in Ref.~\cite{dSC}, we distinguish between the gapped 
phases (e.g., CFL and mCFL phase) and the gapless phases (e.g., gCFL). 
Also,  in order to reflect the physical properties of the mCFL phase,
we suggest to use the term {\em metallic\/} CFL, instead of 
{\em modified\/} CFL as in Ref.~\cite{dSC}. (Note that, in that work, 
the mCFL phase also encompasses the gCFL phase.)

In the case of a large strange quark mass (i.e., the case of $m_s^2/\mu
=320$ MeV shown in Fig.~\ref{m_s-large}), the zero-temperature 
limit corresponds to the gCFL phase. By looking at the corresponding 
temperature dependence of the gap parameters, we see that this case is a
natural generalization of the previous limit of a small strange quark 
mass. There are also three consecutive phase transitions. It is 
noticeable, however, that the separation between the different 
transitions becomes much wider at large $m_s$.

We now take a closer look at the transition between the CFL, the mCFL, 
and the gCFL phase. Let us recall that, at zero temperature, there was 
no symmetry connected with the order parameter, i.e., the number density 
of electrons, that is associated with the CFL $\to$ gCFL phase transition. 
At nonzero temperature, the electron density is not strictly zero in the 
CFL phase as soon as $m_s\neq 0$. Indeed, the arguments of 
Ref.~\cite{enforced} regarding the enforced neutrality of the CFL phase 
do not apply at $T\neq 0$. This leads us to the conclusion that the 
insulator-metal transition between the CFL and the mCFL phase is just 
a smooth crossover at $T\neq 0$. Of course, in principle, we can never 
exclude the existence of a first-order phase transition. Our numerical 
analysis, however, reveals a crossover. The transition can only be 
identified by a rapid increase of the electron density in a relatively 
narrow window of temperatures, see Fig.~\ref{muT}~(a). The location of 
the maximum of the corresponding ``susceptibility'' (i.e., $d n_Q/d T$) 
is then associated with the transition point.

The transition between the mCFL and the gCFL phase corresponds to the 
appearance of gapless quasiparticle modes in the spectrum. We do not 
yet know whether this transition is associated with any physical 
susceptibility. There is no way of telling from the temperature 
dependences in Figs.~\ref{m_s-small}, \ref{m_s-small-zoom} and 
\ref{m_s-large}, whether the corresponding CFL and/or 2SC phases are 
gapless or not. This additional piece of information can only be 
extracted from the behavior of the quasiparticle spectra. We also 
investigated them, but we do not show them explicitly. 

Our results for the phase structure of dense neutral three-flavor
quark matter are summarized in Fig.~\ref{phase-d}. We show the phase 
diagram in the $T$--$m_s^2/\mu$ plane at a fixed value of the quark 
chemical potential, $\mu=500$ MeV, and in the $T$--$\mu$ plane at a 
fixed value of the strange quark mass, $m_s=250$ MeV. The three solid 
lines denote the three phase transitions discussed above. The two dashed 
lines mark the appearance of gapless modes in the mCFL and 2SC phases. 
We could term these as the mCFL $\to$ gCFL and 2SC $\to$ g2SC crossover 
transitions. In addition, as we mentioned above, there is also an 
insulator-metal type transition between the CFL and mCFL phase. 
This is marked by the dotted lines in Fig.~\ref{phase-d}.  

\section{Conclusions}
\label{conclusions}

In this paper, we studied neutral three-flavor quark matter at large 
baryon densities. We obtained a very rich phase structure by varying 
the strange quark mass, the quark chemical potential, and the 
temperature. 

At $T=0$, there are two main possibilities for the strange quark 
matter ground state: the CFL and gCFL phases. These findings confirm 
the results of Ref.~\cite{gCFL} concerning the existence of the 
gapless CFL phase, the estimate of the critical value of the strange 
quark mass $m_s$, and the dependence of the chemical potentials on 
$m_s$. We also confirm that it is the color neutrality condition, 
controlled by the color chemical potential $\mu_8$ which drives the 
transition from the CFL to the gCFL phase \cite{gCFL}. This is in 
contrast to gapless 2SC superconductivity which results from 
electrical neutrality \cite{g2SC}.

Because we use a nine-parameter ansatz for the gap 
function, the results of this paper are more general than those of 
Ref.~\cite{gCFL}. For example, we were able to explicitly study the 
effects of the symmetric pairing channel, described by the sextet-sextet 
gap parameters, that were neglected in Ref.~\cite{gCFL}. As one might 
have expected, these latter modify the quasiparticle dispersion 
relations only slightly. This check was important, however, to see 
that the zero-temperature phase transition from the CFL phase to the 
gapless CFL phase, which is not associated with any symmetry, is robust 
against such a deformation of the quark system.

In this paper, we also studied the temperature dependence of the
gap parameters and the quasiparticle spectra. In particular, 
this study revealed that there exist several different phases of 
neutral three-flavor quark matter that have been predicted in 
the framework of the Ginzburg-Landau-type effective theory in 
Ref.~\cite{dSC}. Our results extend the near-critical 
behavior discussed in Ref.~\cite{dSC} to all temperatures. 
Also, we show how this behavior 
evolves with changing the value of the strange quark mass. The 
only real qualitative difference between our results and the 
results of Ref.~\cite{dSC} is that, instead of the dSC phase, we
find the uSC phase in the phase diagram.

The main result of our paper is the complete phase
diagram of neutral three-flavor quark matter in the $T$--$m_s^2/\mu$
and $T$--$\mu$ plane, shown in Fig.~\ref{phase-d}. In this figure,
all symmetry related phase transitions are denoted by solid lines.
(The symmetries of all phases appearing in this figure were discussed
in Ref.~\cite{dSC}.) In the mean-field approximation used here, all 
of these transitions are second-order phase transitions. After 
taking into account various types of fluctuations, the nature of 
some of them may change \cite{fluct}. A detailed study of this issue is, 
however, outside the scope of this paper. The dashed lines in 
Fig.~\ref{phase-d} separate the mCFL and regular 2SC phases 
from the gapless CFL and gapless 2SC phases. These cannot be real 
phase transitions, but are at most smooth crossovers. 
At $T=0$, there is an insulator-metal phase transition between the 
CFL and the gCFL phase \cite{gCFL}. At nonzero temperature, there 
exists a similar insulator-metal type transition between the 
CFL and the mCFL phase, given by the dotted lines in Fig.~\ref{phase-d}.

\begin{acknowledgments}
The authors would like to thank Michael Buballa and Mei Huang 
for interesting discussions, and Axel Maas for suggesting
to include Fig.~\ref{phase-d}~(b). The work of I.S. was supported by 
Gesellschaft f\"{u}r Schwerionenforschung (GSI) and by 
Bundesministerium f\"{u}r Bildung und Forschung (BMBF).
\end{acknowledgments}



\begin{widetext}

\begin{figure}
    \includegraphics[width=0.9\textwidth]{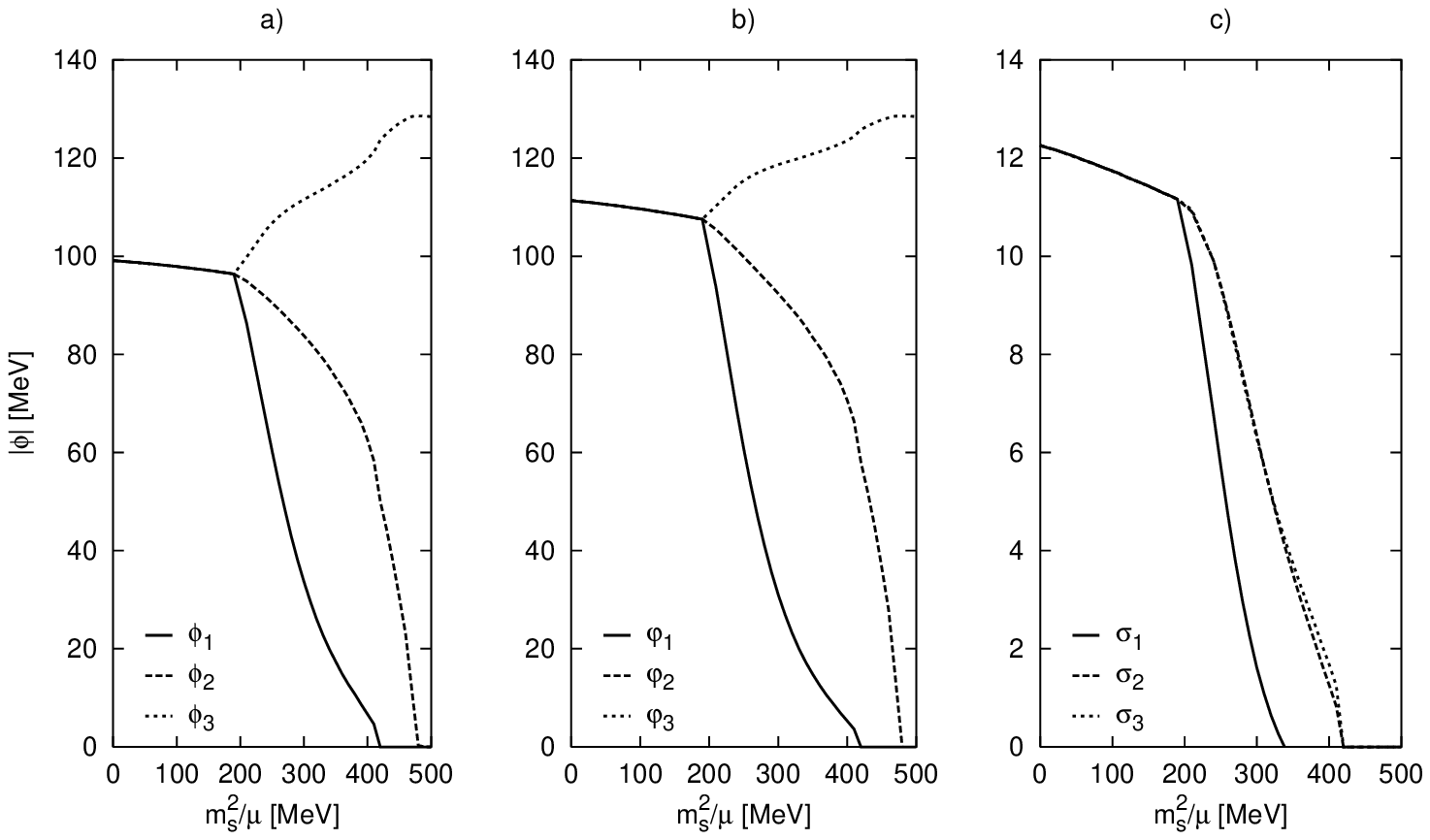}
    \caption{The absolute values of the gap parameters as a function 
       of $m_s^2/\mu$ for neutral color-superconducting quark matter 
       at $T=0$ and $\mu=500$ MeV. The actual values of the gap parameters 
       shown in panel a) are negative.}
    \label{phi2ms}
\end{figure}

\begin{figure}
    \includegraphics[width=0.65\textwidth]{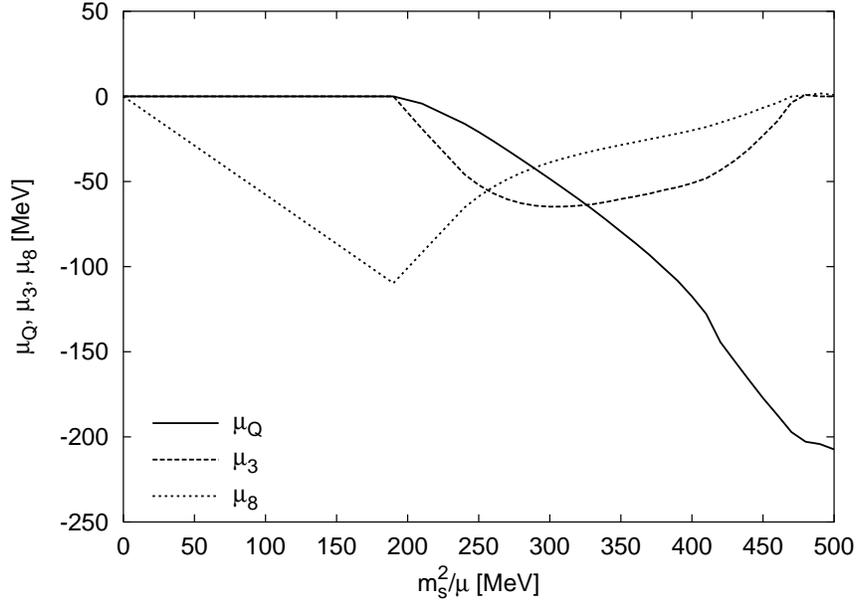}
    \caption{The electrical and color chemical potentials as a function 
       of $m_s^2/\mu$ of electrical and color neutral color-superconducting 
       quark matter at $T=0$ and $\mu=500$ MeV.}
    \label{ch2ms}
\end{figure}

\begin{figure}
    \includegraphics[width=0.9\textwidth]{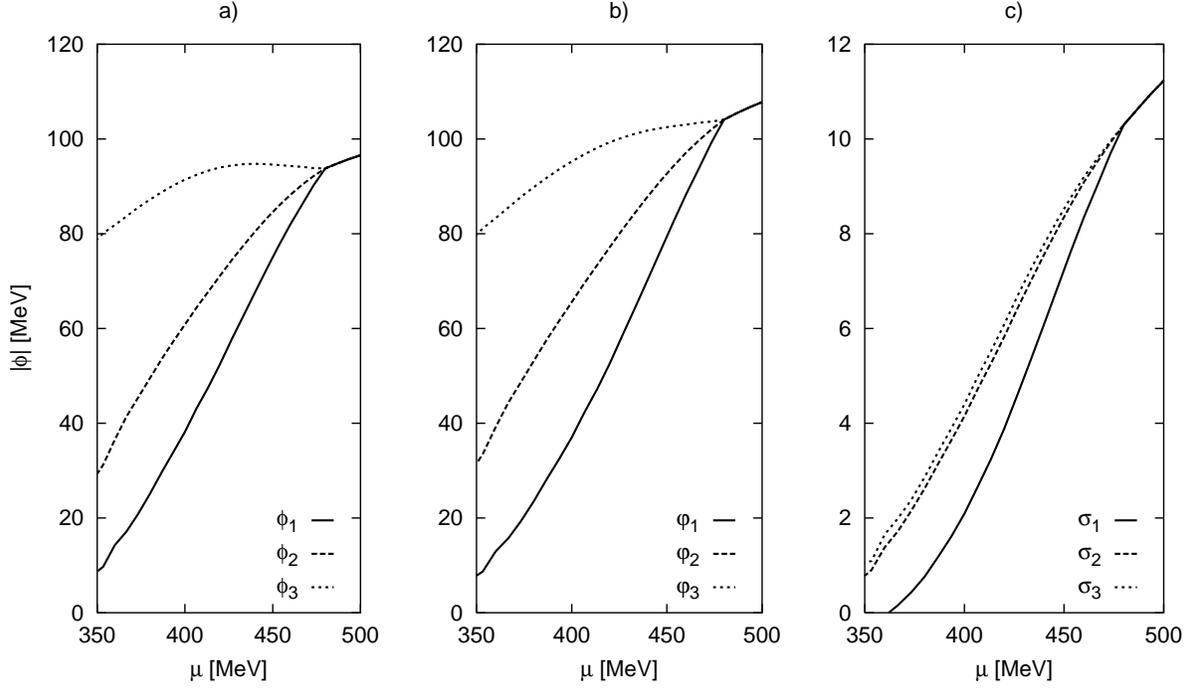}
    \caption{The absolute values of the gap parameters as a function of 
      $\mu$ of electrical and color neutral color-superconducting quark 
      matter at $T=0$ and $m_s=300$ MeV. The actual values of the gap 
      parameters shown in panel a) are negative.}
    \label{phi2mu}
\end{figure}

\begin{figure}
    \includegraphics[width=0.9\textwidth]{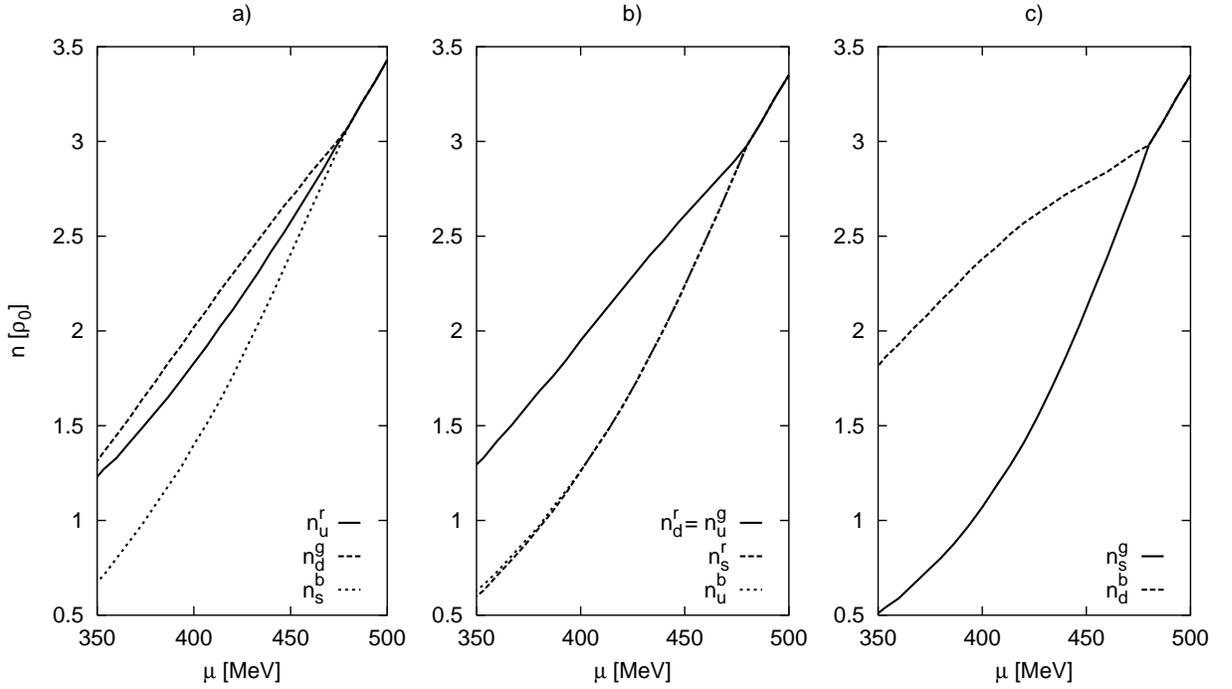}
    \caption{The number densities of each quark color and flavor as a function 
      of $\mu$ for electric and color neutral color-superconducting quark 
      matter at $T=0$ and $m_s=300$ MeV. The densities are given in units 
      of the saturation density of nuclear matter, $\rho_0=0.15$~fm$^{-3}$.}
    \label{n2mu}
\end{figure}

\begin{figure}
    \includegraphics[width=0.875\textwidth]{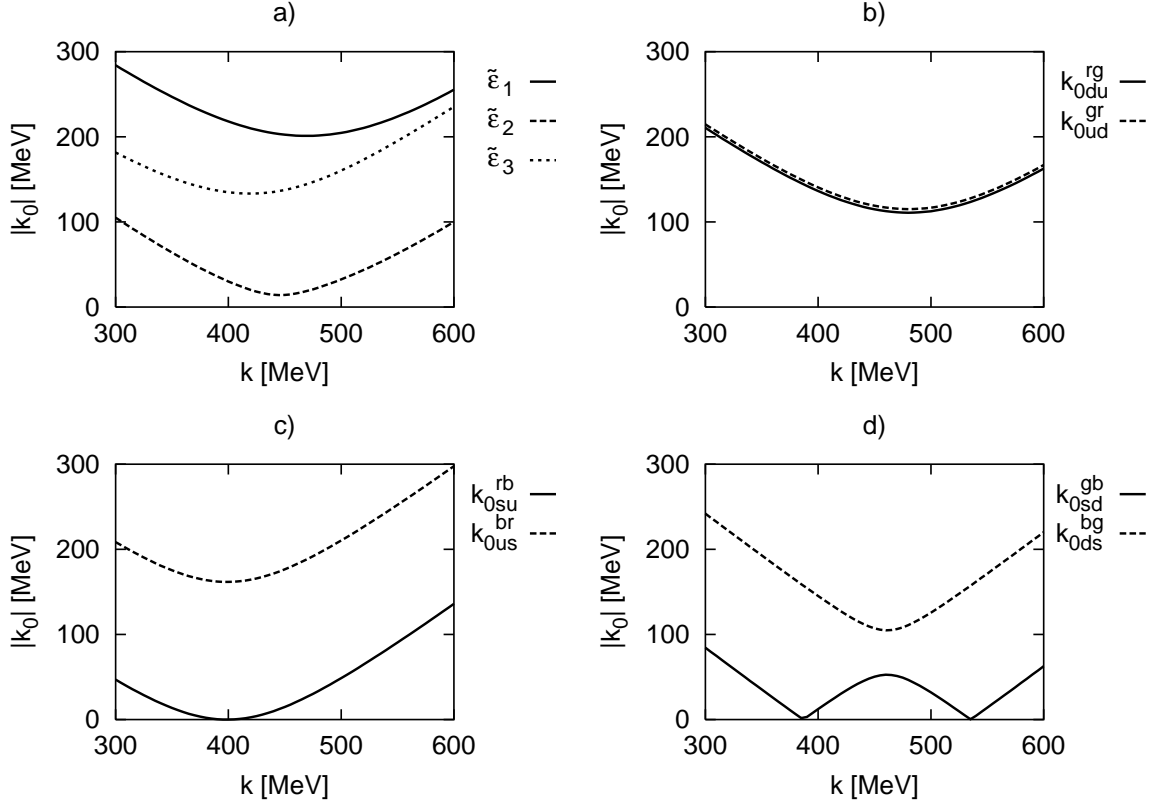}
    \caption{The quasiparticle dispersion relations for electrical and 
    color neutral color-superconducting quark matter at $T=0$, $\mu=500$ 
    MeV, and $m_s=400$ MeV.}
    \label{disp}
\end{figure}

\begin{figure}
    \includegraphics[width=0.9\textwidth]{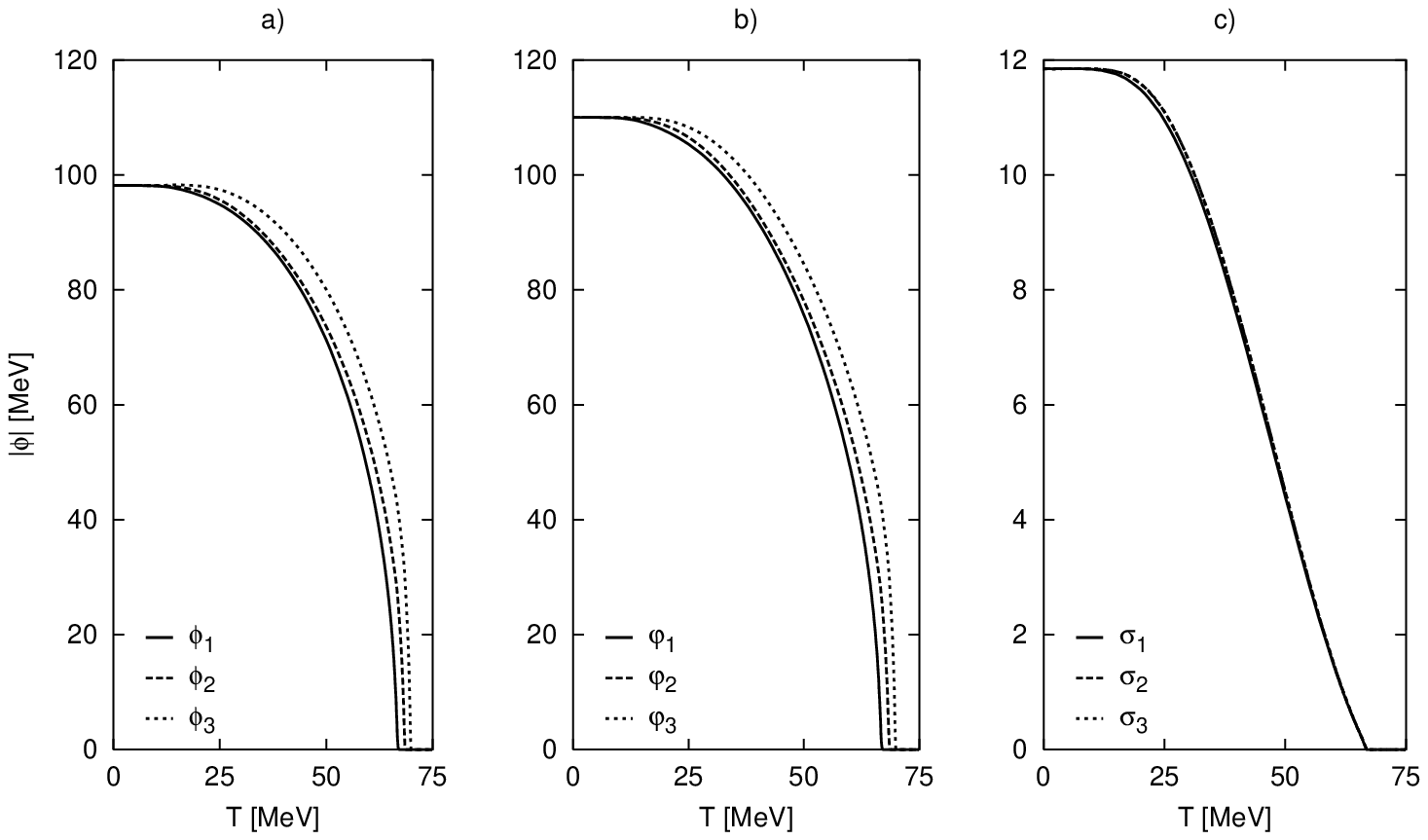}
    \caption{The temperature dependence of the gaps in the case 
       of a small strange quark mass, $m_s^2/\mu=80$ MeV. Note that the 
       actual values of the gap parameters shown in panel a) are negative.}
    \label{m_s-small}
\end{figure}

\begin{figure}
    \includegraphics[width=0.9\textwidth]{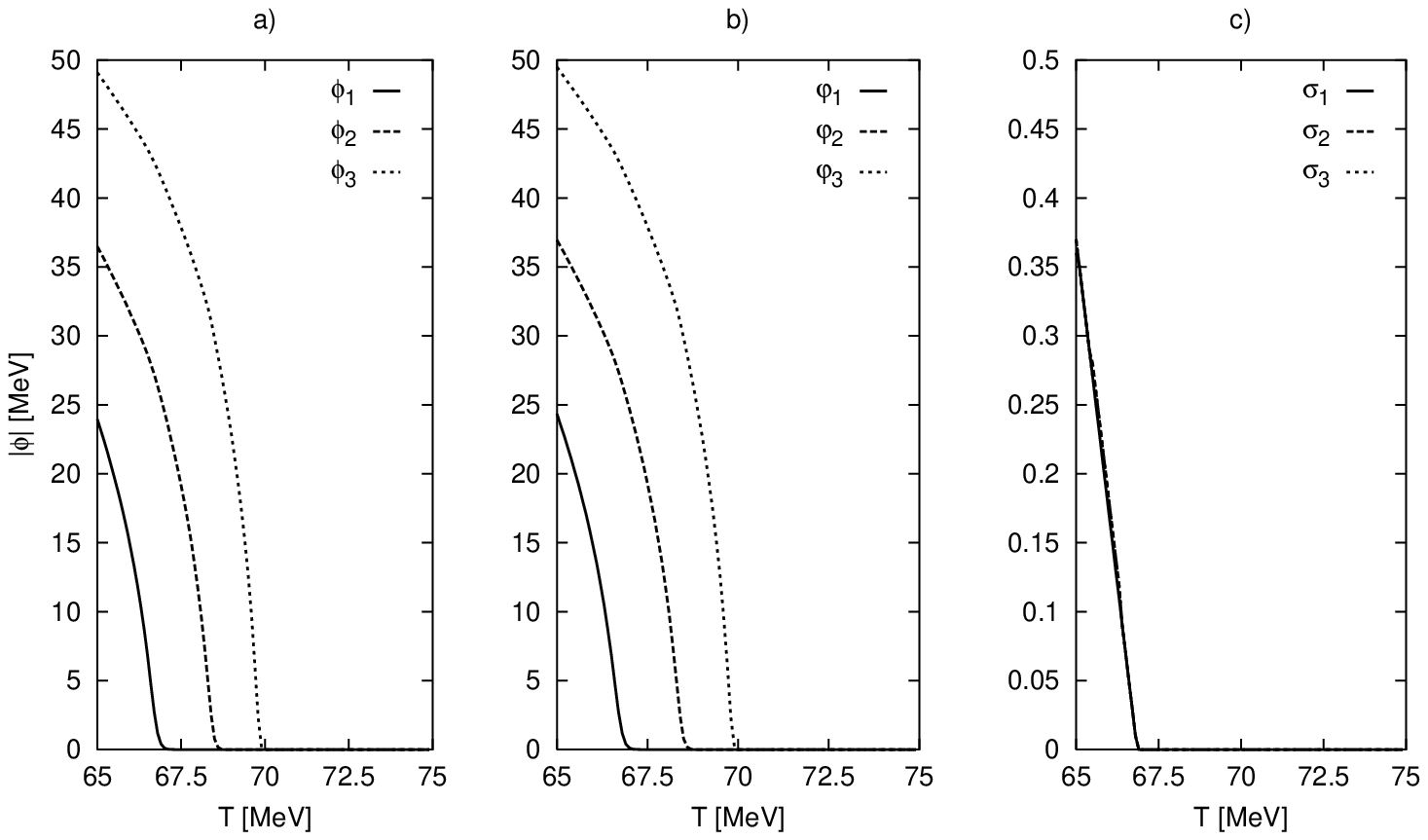}
    \caption{The near-critical temperature dependence of the gaps 
    in the case of a small strange quark mass, $m_s^2/\mu=80$ MeV.
    Note that the actual values of the gap parameters shown in panel a) 
    are negative.}
    \label{m_s-small-zoom}
\end{figure}

\begin{figure}
    \includegraphics[width=0.9\textwidth]{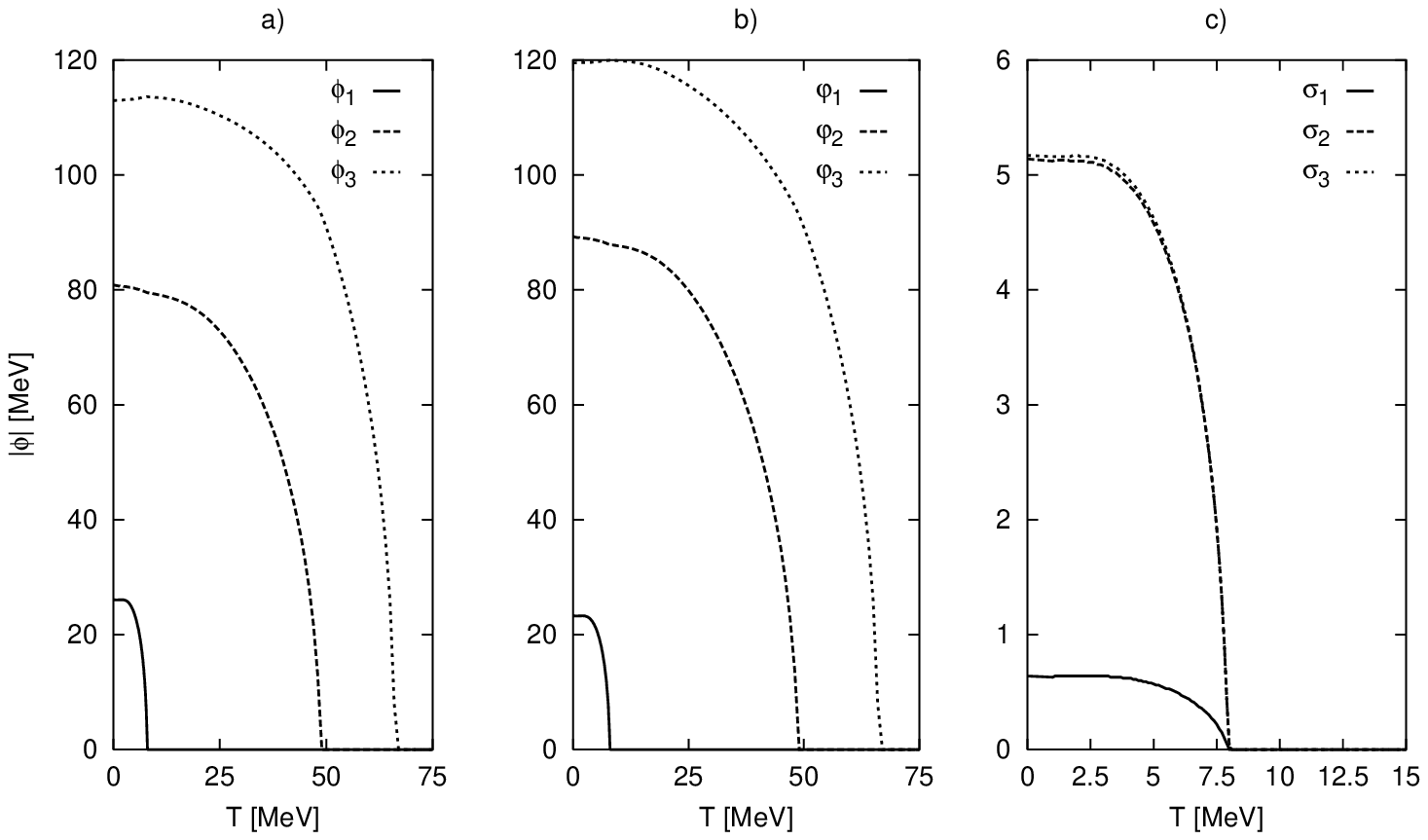}
    \caption{The temperature dependence of the gaps in the case of
    a large strange quark mass, $m_s^2/\mu=320$ MeV. Note that the 
    actual values of the gap parameters shown in panel a) are negative.}
    \label{m_s-large}
\end{figure}

\begin{figure}
    \includegraphics[width=0.9\textwidth]{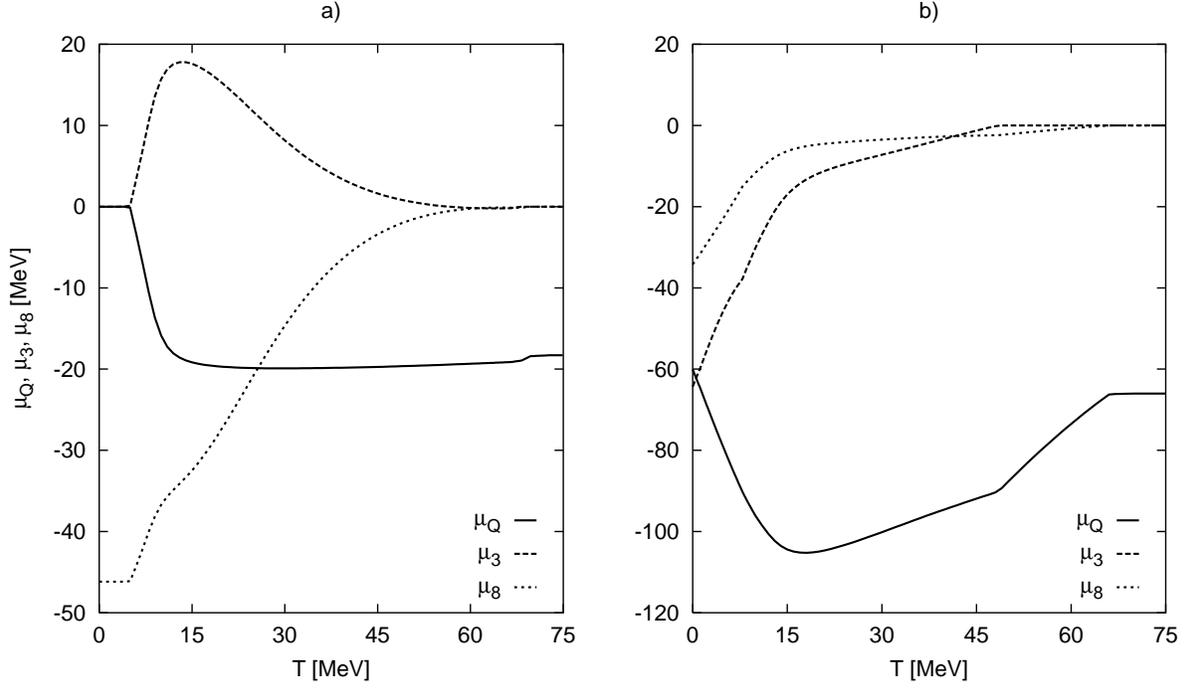}
    \caption{The temperature dependence of the electrical and color 
    chemical potentials for $m_s^2/\mu=80$ MeV (left panel) and 
    for $m_s^2/\mu=320$ MeV (right panel). The quark chemical potential 
    is taken to be $\mu=500$~MeV.}
    \label{muT}
\end{figure}

\begin{figure}
\includegraphics[width=0.9\textwidth]{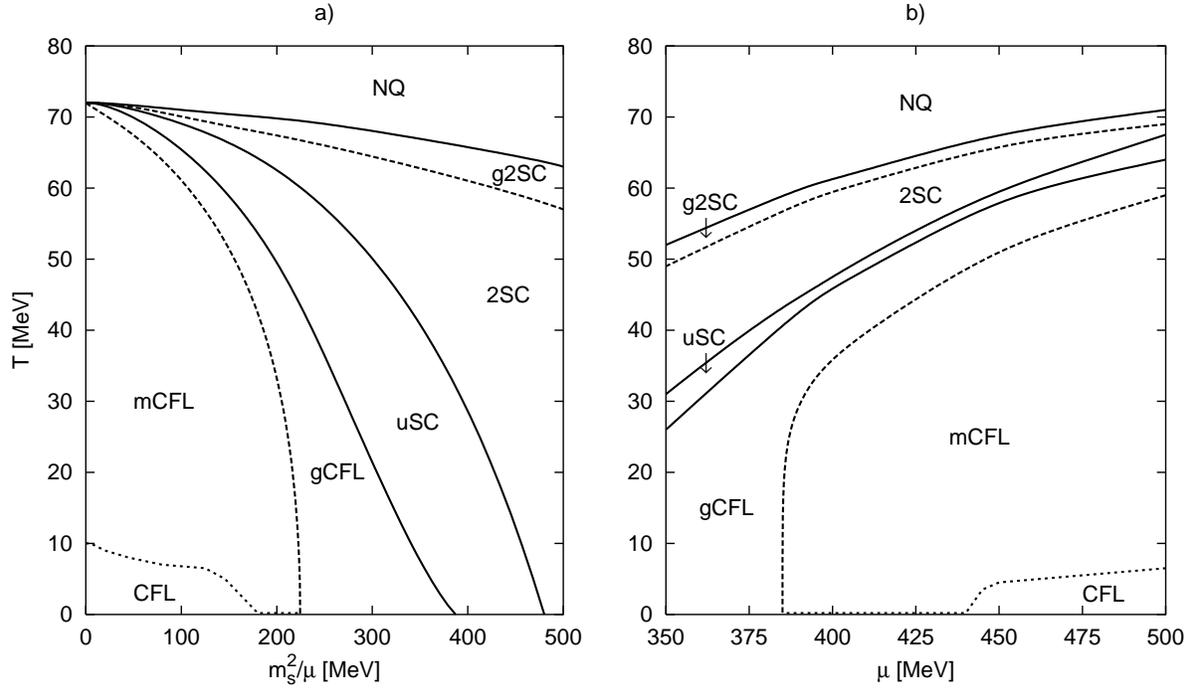}
\caption{The  phase diagram of neutral three-flavor quark 
     matter in the plane of temperature and $m_s^2/\mu$ (left panel)
     and in the plane of temperature and quark chemical potential (right 
     panel). The results in the left panel are for a fixed value 
     of the quark chemical potential, $\mu=500$ MeV. The results in 
     the right panel are for a fixed value of the strange quark 
     mass, $m_s=250$ MeV. The dashed lines are associated with the 
     appearance of additional gapless modes in the spectra. The dotted 
     lines indicate the insulator-metal crossover.}
    \label{phase-d}
\end{figure}

\end{widetext}
\end{document}